\documentclass{PoS}

\title{ Time evolution of a medium-modified jet}

\ShortTitle{Time evolution of a medium-modified jet}

\author{\speaker{Liliana Apolinário}\\
    LIP\\
    Av. Prof. Gama Pinto, n.2	Complexo Interdisciplinar (3is)	1649-003 Lisboa Portugal
    E-mail: \email{liliana@lip.pt}}

\abstract{The presence of a hot and dense medium, produced in ultra-relativistic heavy-ion collisions, is known to modify the parton shower evolution. Several observations of the resulting intra-jet activity show significant modifications of what can be considered as a medium-modified jet from a "vacuum" (proton-proton) reference. These modifications, generically known as jet quenching effects, are the result of the multiple interactions of the parton shower with the produced fast-evolving quark-gluon plasma (QGP). Recent efforts have tried to assess the time dependence of jet quenching effects, with particular focus on late or early dynamics. In this talk, we show a novel tool that can potentially evaluate the full-time evolution of the QGP by applying jet reclustering techniques. The result can bring novel insights into the QGP expansion as well as shed some light on how to re-sum, in a consistent way, vacuum-like and medium-like emissions into a single parton shower evolution equation.}

\FullConference{
European Physical Society Conference on High Energy Physics - EPS-HEP2019 -\\
			10-17 July, 2019\\
			Ghent, Belgium}

\begin{document}

\section{Introduction}

In ultrarelativistic heavy-ion collisions, extreme temperatures and densities allow the production of a hot and dense medium of quarks and gluons, called the Quark-Gluon Plasma (QGP). With an initial volume proportional to the overlap area of the incoming nuclei, this state of matter expands very fast, dropping its temperature bellow the deconfinement transition (or cross-over), forming a high multiplicity event. Due to its short lifetime, its study is made only indirectly by analysing the final particles that are the result of the QGP production and expansion (generically known as soft probes). Alternatively, one can use high momentum particles that are produced simultaneously within the collision (hard probes). Since they propagate through the produced medium, they will suffer modifications (generically known as \textit{jet quenching}) induced by their interactions with the QGP. By comparing these observations to a reference (usually taken as proton-proton collisions), it is possible to infer the properties of the medium that these probes travelled through. 

The Relativistic Heavy-Ion Collider (RHIC) and the Large Hadron Collider (LHC), collected a paramount of results in both sectors. These include hydrodynamic flow patterns, suppression of heavy-quark bound states, hadrochemistry of the final state, and modifications of the fragmentation of energetic partons that traverse the medium (see \cite{Foka:2016vta,Foka:2016zdb} for an overview of the results). Together, they brought a unique insight into the properties of the QGP (see \cite{Marco,Dennis,JanFiete,Carlos}). However, a limiting factor of all these observables is that they are the integrated result of a fast-evolving and extended medium. As such, there is a strong time-dependence of the medium properties that require novel strategies to measure the time-structure of the QGP.

\section{Probing the QGP time evolution}

Attempts to probe the space-time evolution of the QGP have been put forward recently. In particular, the works by \cite{Andres:2016iys} and \cite{Apolinario:2017sob}. 

In the former, using a framework to change energy loss effects during early the stages of the parton shower development, the authors show that the nuclear modification factor, $R_{AA}$ seems to be insensitive to a "switching off" of quenching effects during the first 0.6~fm/c. Conversely, the high momentum anisotropic flow appears to be quite sensitive to such change. In particular, the scenario in which the initialisation time for energy loss effects starts after 0.6~fm/c seems to be preferred by both ATLAS and CMS results. This observation shows the potential to constrain the dynamics of the initial stages of medium evolution. 

In the later, there is a proposal to use semi-leptonic decay of $t\bar{t}$ events as a tomographic probe of the QGP. In particular, this constitutes a time-delayed probe of the medium. The (colourless) $W$-boson decays into a $q\bar{q}$ pair, and by selecting higher momentum (and so highly boosted) particles, one can trigger on reconstructed $W$-bosons that are produced at later times. Its hadronic products will interact with the resulting QGP that is, by then, already in a diluted state. The result will be a shift in the reconstructed $W$-boson mass that will depend on the reconstructed top transverse momentum, $p_{T,top}^{reco}$. In particular, the expected evolution from low to high $p_{T,top}^{reco}$, is from a scenario in which particles start to interact with the medium immediately after its production (fully quenching scenario) towards vacuum expectations (proton-proton). If the former is known (from, for instance, boson-jet energy imbalance), by measuring the reconstructed $W$-boson mass as a function of the top transverse momentum, it should be possible to perform a full tomographic analysis of the QGP time evolution. While this result shows that in the Future Circular Collider, a scan of the entire QGP lifetime is possible, currently, even with the HE-LHC upgrade, there is limited discrimination between short and long-lived mediums.


While there are possible exploration avenues to constrain initial and later times of the QGP evolution, it would be desirable to have a probing tool that would allow direct sensitivity to the different QGP timescales. In the next section, I will try to argue that among our current probes, jets might be a good candidate for such studies.

\section{Jets: a multi-scale probe}

Jet production is abundant at both RHIC and the LHC. They are a space-time evolving structure that is the result of the QCD parton shower initiated by a high momentum object. These successive multiple parton emission process occurs from a high energy scale, until a lower energy scale (when non-perturbative processes become relevant).

In a perturbative formulation, such iterative process is, at leading order, dominated by angular ordered emissions, such that:
\begin{equation}
dN^{\omega \rightarrow 0} \sim \alpha_s C_F \frac{d\omega}{\omega} \frac{\sin \theta d\theta}{1-\cos \theta} \Theta( \cos \theta - \cos \theta_{q\bar{q}}) 
\end{equation}
describes the spectrum of radiated soft gluons, with energy $\omega$ and angle $\theta$, off a quark that was previously emitted with angle $\theta_{q\bar{q}}$. 

The time for a given emission to be regarded as an independent source of emission can be estimated as the lifetime of the virtual quark-gluon state. In the centre-of-mass reference frame of the virtual pair, this time goes as:
\begin{equation}
	\Delta t \sim \frac{1}{\Delta E} \sim \frac{1}{m_{virtual}} \, .
\end{equation}
In the laboratory frame, due to the Lorentz boost, this parametric estimate yields:
\begin{equation}
	\tau_{form} \simeq \frac{1}{z (1-z) E \theta^2} \, ,
\end{equation}
where $E$ is the energy of the parent quark, $z$ the energy fraction carried by the gluon and $\theta$ the angle between the final quark and gluon. As such, the full parton shower should encode all such parametric estimates for lifetimes in a somehow ordered way (at leading order).

\subsection{Vacuum vs in-medium parton shower}

In vacuum, the Monte Carlo approach of iterative parton emissions does not have a physical realisation. The probabilistic interpretation of quantum emissions, driven by an ordering variable (virtuality, transverse momentum, or angle), is known to capture the final picture of a jet, i.e, the set of final particles. Because the whole transformation process of a single parton to a collection of final state particles takes place without an intermediate measure of the process, it is not known if the iterative process of the Monte Carlo approach had that particular realisation. 
However, if the same process now takes place inside a medium whose density is evolving with time, it might be possible to relate the characteristics of a given parton emission with the medium density that participated in the emission of that given particle. From there, one should be able to capture the concept of a physical time in the parton shower development.

As a first study along this line, it was used a \textit{vacuum} Monte Carlo event generator (PYTHIA8.2.40 \cite{Sj_strand_2006}) and a Monte Carlo event generator that includes medium-induced effects to simulate how the jet is modified by the presence of a QGP (JEWEL 2.2.0 \cite{Zapp:2013vla}). As a first study, and as a proof of concept, medium-recoil effects were not included to facilitate the tracking/identification of particles that come exclusively from the parton shower. Future studies will naturally include these effects. 

For the results presented in this manuscript, it was simulated di-jets events at $5.02$TeV at hadron level ([0-10]\% centrality class within JEWEL and pre-defined simple medium parameters). The list of final jets is obtained using the anti-$k_t$ algorithm\cite{Cacciari_2008}, with radius $R = 0.5$, a minimum transverse momentum of $p_{T,min}= 100$ GeV and a pseudo-rapidity window of $|\eta| < 1$, performed within the FastJet package\cite{Cacciari_2012}. The choice of such a narrow window is to reduce effects that are not Lorentz invariant. In order to have already a \textit{time} structure within the jet, we decided to recluster the jets with the generalized $k_t$ jet algorithm:
\begin{equation}
	d_{ij} = \min (p_{t,i}^{2p}, p_{t,j}^{2p}) \frac{\Delta R_{ij}^2}{R^2} \ , \ d_{i,B} = p_{t,i}^{2p} \, ,
\end{equation}
with $p = 0.5$. This value provides the same parametric estimate that the inverse of $\tau_{form}$:
\begin{equation}
	d_{ij} \sim p_{T} \theta^2 \sim \tau_{form}^{-1} \, .
\end{equation}
While this choice might appear as unorthodox for a vacuum parton shower, we are already anticipating the physics of a medium-modified parton shower. JEWEL already includes some formation time effects for subsequent emissions. In particular, the interplay between competing radiative processes is governed by the formation time (the emission with a longer formation time is discarded). For the vacuum parton shower, since $\tau_{form}$, as it is defined, depends intrinsically on the kinematical properties of the radiated quanta (such as emission angle and transverse momentum), we expect that this reclustering also captures QCD-inspired splittings, somehow ordered from long to short $\tau_{form}$. Naturally, other reclustering algorithms (such as Cambridge-Aachen\cite{Dokshitzer:1997in,Wobisch:1998wt}), might also be investigated.

\subsection{Jet Splitting function}

\begin{figure}
\centering
 \includegraphics[width=0.6\linewidth]{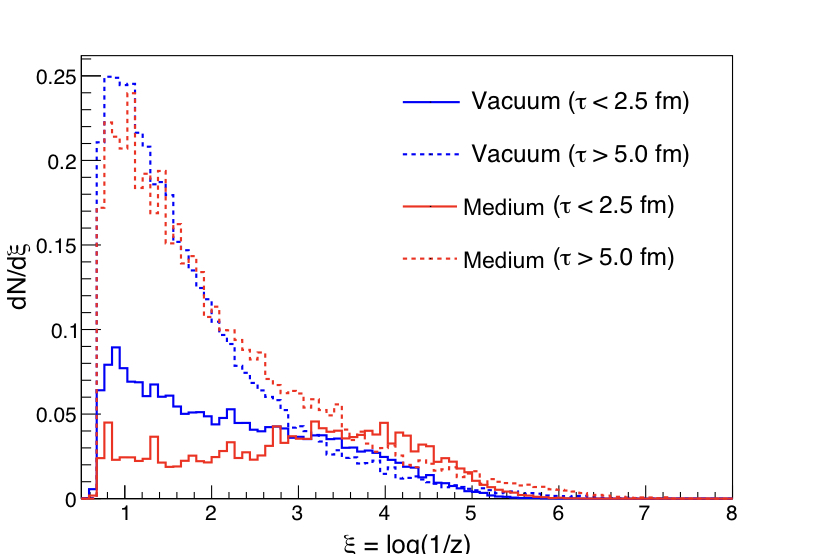}
 \caption{Fragmentation function of the jet recursive splittings obtained from the primary branch when using a generalised $k_t$ algorithm with $p = 1$. The $z$ is the energy fraction of the sub-leading sub-jet. The label \textit{Vacuum} corresponds to PYTHIA8 and \textit{Medium} to JEWEL2.2.0. Two intervals of formation time are also shown: $\tau_{form} = \tau < 2.5$ fm/c and $\tau_{form} = \tau > 5.0$ fm/c.}
 \label{fig:plot1}
\end{figure}

With the procedure from the last section, and selecting only the leading jet, we recursively identify the sub-jet prongs from the primary branch by unclustering the jet sequence. Two different $\tau_{form}$ intervals were selected: $\tau_{form} < 2.5$~fm/c and $\tau_{form} > 5$~fm/c. These values were chosen to select splittings that took place inside the medium and splittings that were, in its most, driven by vacuum physics (or in a state where the QGP is already very diluted). The corresponding fragmentation function of the sub-jets that fall into each window is shown in figure \ref{fig:plot1}. In this plot, $\xi = \log (1/z)$, with $z$ being the fraction of energy carried by the sub-leading jet. The solid corresponds to the interval of small $\tau_{form}$ while the dashed to the splittings with a formation time that falls into the window of $\tau_{form} > 5$~fm/c. While the later does now show any significant modification between JEWEL (in red, labelled medium) and PYTHIA (in blue, labelled vacuum), we see a clear modification of this distribution when we move to sub-jets that were formed early, within the QGP lifetime. 

\begin{figure}
\centering
 \includegraphics[width=0.6\linewidth]{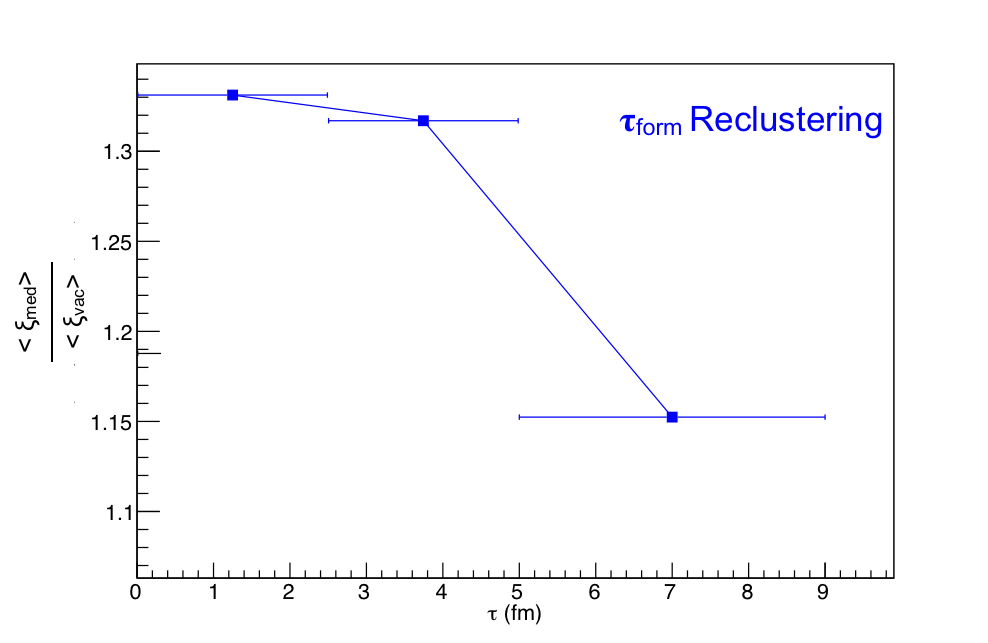}
 \caption{Medium-to-Vacuum ratio of the average value of $\xi$ (to be read from figure \ref{fig:plot1}), as a function of formation time.}
 \label{fig:plot2}
\end{figure}

As a simple exercise to verify that the characteristics of the radiation could be potentially related to the time evolution of the QGP, we took the average value of the above distribution, $\left\langle \xi \right\rangle$, and plotted the ratio medium over vacuum as a function of the obtained formation time. This is shown in figure \ref{fig:plot2}, for 3 intervals of $\tau_{form}$: [0-2.5]~fm/c; [2.5 - 5]~fm/c; [5-9]~fm/c. There is clear linear depletion that is approaching to $1$, as expected, suggesting the dilution of the QGP density that is incorporated within JEWEL (a Bjorken expansion). Naturally, the units in $x$-axis are to be taken as parametric estimates. Moreover, the $\tau_{form}$ itself is obtained using a generalised $k_t$ re-clustering. Being so, the kinematic information that is retrieved from each of the sub-jets in the process of successive unclustering is also affected by this choice.  

\begin{figure}
\centering
 \includegraphics[width=0.6\linewidth]{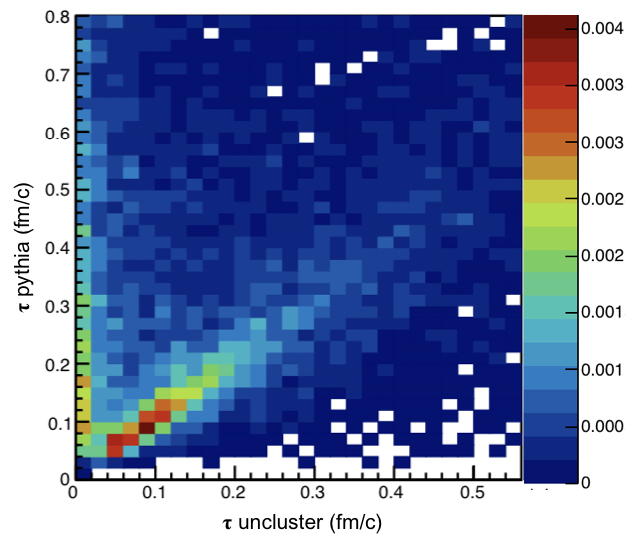}
 \caption{Correlation between the formation time obtained from the jet unclustering method ($x$-axis) and the PYTHIA8 parton shower history ($y$-axis).}
 \label{fig:plot3}
\end{figure}

With this concern in mind, it was evaluated how well the reclustered sub-jets would correspond to the emissions in the Monte Carlo history. Again, caution should be taken when interpreting the obtained results as the Monte Carlo history is not meant to reproduce the exact physical process in each quantum emission, but the final result of the whole parton shower. Moreover, the iterative soft drop process is known to yield a jet splitting function ($z_g$)\cite{Sirunyan:2017bsd,Acharya:2019djg} that is comparable to the QCD splitting functions in proton-proton collisions. While the same measurement has been reported in heavy-ions, the link between this measurement and that to a QCD-like splitting is still unclear\cite{Milhano:2017nzm}. However, one can take this exercise to understand if QCD-like emissions are tagged in the same way when using this reclustering algorithm ($p = 1$). In this part, we took only PYTHIA and the corresponding correlation plot between the formation time that is computed within the Monte Carlo event generator ($y$-axis) and the un-clustering method ($x$-axis) is shown in figure \ref{fig:plot3}. There is indeed a direct correlation between the two quantities, on top of some wrong branch identifications that spread along the values of $\tau_{form}$. Another key feature of this figure is a vertical line for shorter formation times when doing the un-clustering. These splittings correspond to large angle radiations that will not be part of the jet as they will end in a region of $(\eta,\phi)$ phase space at an $R > 0.5$ from the jet axis.

\section{Conclusions}

One of the goals of the heavy-ion program in ultra-relativistic heavy-ion collisions is to measure the QGP properties accurately. Since the produced medium is fast expanding, aiming for precision measurements requires an evaluation of the time-dependence of the medium characteristics. Therefore, there is an urgent need to evaluate the QGP space-time evolution. 

In this work, we made a first attempt to use jet substructure as a proxy for the time evolution of the QGP. The results show that selecting jet prongs with different formation times can indeed be used to quantify the effect of the QGP on those \textit{splittings}. In particular, prongs with longer formation have a fragmentation function that is similar when comparing PYTHIA and JEWEL, while sub-jets with shorter formation times show distributions that are heavily modified. This can provide an exploration avenue to measure the time density profile of the medium. While the results seem promising, there are several challenges ahead. The matching between the parton shower branch (from the Monte Carlo event generator) and that to the reconstructed sub-jet was investigated within PYTHIA yielding moderate results\footnote{At the time of the writing of these proceedings, there was a dedicated workshop to investigate the space-time structure of jet quenching\cite{GSI}. Results accomplished during the meeting showed that jet grooming techniques could significantly improve the matching history between Monte Carlo event generators and the jet reclustering algorithm.}. Further investigation is required to check such universality across different quenching models. This is under investigation, and the results will be released in a future manuscript. 

\textbf{Acknowledgements:} The author would like to thank to R. Conceição, G. Milhano and J. Thaller for useful discussions. This work was supported by the Portuguese Funda\c{c}\~{a}o para a Ci\^{e}ncia e Tecnologia under contracts DL57/2016/CP1345/CT0004 and CERN/FIS-PAR/0022/2017.

\bibliography{bibliography}{}
\bibliographystyle{JHEP}

\end{document}